\documentclass[showpacs]{revtex4}
\usepackage{indentfirst}
\usepackage[mathscr]{eucal}
\usepackage[dvips]{graphicx}
\usepackage{amsmath}
\def\lsim{\mathrel{\raise2pt\hbox to 8pt{\raise -5pt\hbox{$\sim$}\hss{$<$}}}}
\def\bmath#1{\mbox{\boldmath $#1$}}
\def\ds{\displaystyle}
\def\rb#1{\raisebox{1.5ex}[0pt]{#1}}
\begin{document}
\title{Tensor coupling and pseudospin symmetry in nuclei}
\author{P. Alberto}
\affiliation{Departamento de F\'{\i}sica and Centro de F\'{\i}sica
Computacional, Universidade de Coimbra, P-3004-516 Coimbra,
Portugal}
\author{R. Lisboa}
\affiliation{Instituto de F\'{\i}sica, Universidade Federal
Fluminense, 24210-340 Niter\'oi, Brazil}
\author{M. Malheiro}
\affiliation{Instituto de F\'{\i}sica, Universidade Federal
Fluminense, 24210-340 Niter\'oi, Brazil \\and Departamento de F\'{\i}sica, 
Instituto Tecnol\'ogico de Aeron\'autica, CTA,
12228-900, S\~ao Jos\'e dos Campos, SP, Brazil}
\author{A. S. de Castro}
\affiliation{Departamento de F\'{\i}sica e Qu\'{\i}mica, Universidade Estadual Paulista, 12516-410 Guaratinguet\'a, SP, Brazil}
\date{\today}


\begin{abstract}
\noindent In this work we study the contribution of the isoscalar tensor
coupling to the realization of pseudospin
symmetry in nuclei. Using realistic values for the tensor coupling
strength, we show that this coupling reduces noticeably the pseudospin splittings, 
especially for single-particle
levels near the Fermi surface. By using an energy decomposition of
the pseudospin energy splittings, we
show that the changes in these splittings come by mainly through
the changes induced in the lower radial wave function for the low-lying pseudospin
partners, and by changes in the expectation value of the 
pseudospin-orbit coupling term for surface partners.
This allows us to confirm the conclusion already reached in previous studies,
namely that the pseudospin symmetry in nuclei is of a dynamical nature.

\end{abstract}

\maketitle


\section{Introduction}
\label{Sec:Introduction}

Pseudospin symmetry is a concept that appeared in nuclear physics
more than 30 years ago \cite{npa137,plb30} to
account for the observation, in heavy nuclei, of the
quasidegeneracy of orbitals with quantum numbers
$(n,\,l,\,j=l+1/2)$ and $(n-1,\,l+2,\,j=l+3/2)$ (for fixed $n$
and $l$). Such pairs of single-particle states are known as pseudospin
partners.
This doublet structure is related to the
pseudo-orbital angular momentum and pseudospin quantum numbers $\tilde{l}%
=l+1$ and $\tilde{s}=s=1/2$ \cite{prl68,prl74,prc58,prc59,gino},
respectively. The former, as noted by Ginocchio \cite{gino}, is
just the orbital angular momentum of the lower component of the
Dirac spinor. Pseudospin partners are doublets with
the same $\tilde{l}$. For example, for the partners $%
[ns_{1/2},(n-1)d_{3/2}]$, $\tilde{l}=1$, and for
$[np_{3/2},(n-1)f_{5/2}]$ one has $\tilde{l}=2$.

The existence of degenerate or quasi-degenerate pseudospin
partners is connected to a SU(2) symmetry of the Dirac equation
with only scalar $S$ and vector $V$ potentials such that $V=-S$,
regardless of the particular shapes of these potentials
\cite{bell}. It happens that in the Relativistic Mean-Field
Theories of Nuclei (RMF) \cite{wal,serewal,prc41}, the sum
$\Sigma=S+V$ is small at the nuclear energy scale, so that this
symmetry, known as pseudospin symmetry, provides a natural
explanation of the existence of quasi-degenerate pseudospin
partners in nuclei referred to before
\cite{gino,plb245,prl87,pmmdm1}. However, since in RMF theories
$\Sigma$ acts as binding potential for the nucleons, it is not
possible to have exact pseudospin symmetry in nuclei. Further
works have shown that the particular shape of $\Sigma$, not just
its smallness, can affect the pseudospin energy splittings and
also explain the isospin dependence of these splittings
\cite{pmmdm1,pmmdm2,prc67,bjp34}. Moreover, using an energy
decomposition coming from the Schr\"odinger-like equation for the
lower component of the Dirac spinor, it was shown that the
observed pseudospin splitting arises from a cancellation of the
several energy components, and not primarily from the
pseudospin-orbit term, which is proportional to the derivative of
$\Sigma$ \cite{pmmdm2,prc67}. Altogether, this led us to conclude
that, in nuclei, pseudospin symmetry is realized in a dynamical
way. A similar conclusion was reached by Marcos {\it et al.}
\cite{marcos,marcos2}.

The tensor coupling has been used in studies of nuclear properties with effective Lagrangians
including RMF theories by Furnstahl {\it et al.} in Ref.~\cite{furnstahl}, and in the relativistic Hartree approach model,
studied by Mao in Ref.~\cite{Mao}. Those works assessed its influence on nuclear observables,
namely the spin-orbit splitting of single-particle levels in nuclei, the result being that
the tensor coupling, a higher order term in a relativistic expansion,
increases significantly the spin-orbit coupling.
This suggests that the tensor coupling could have
a significant contribution to pseudospin splittings in nuclei as well. This contribution
is expected to be particularly relevant for the levels near the Fermi surface,
since the tensor coupling depends on the derivative of a vector
potential, which has a peak near the Fermi surface for typical nuclear mean-field vector potentials.

The tensor coupling has also been used as a natural way to
introduce the harmonic oscillator in a relativistic (Dirac)
formalism. In a recent paper, it was shown that the harmonic
oscillator with scalar and vector potentials can exhibit an exact
pseudospin symmetry \cite{prc69,gino_oh}. When this symmetry
is broken ($\Sigma\neq 0$), the breaking term is quite large,
manifesting its nonperturbative behavior. However, if a tensor
coupling is introduced, the form of harmonic-oscillator potential
can still be maintained with $\Sigma=0$, but the pseudospin
symmetry is broken perturbatively \cite{ronai2}.

The tensor interaction has also been considered in order to explain how the
spin-orbit term can be small for $\Lambda$-nucleus and large in
the nucleon-nucleus case \cite{chiappa}. It is assumed that in the
strange sector (case of $\Lambda$) the tensor coupling is large
and the spin-orbit term obtained from this interaction can cancel
in part the contribution coming from the scalar and vector
interactions. This result shows that the tensor interaction can
change strongly the spin-orbit term. In this spirit, we want to
investigate if this interaction can also affect the
pseudospin-orbit term (spin-orbit of the lower component)
\cite{prc58,prc59,pmmdm2}. We address this problem by performing a mean-field
calculation for the neutron levels of $^{208}$Pb, using mean-field
Lorentz vector and scalar potentials with a Woods-Saxon shape.
These potentials were used in the previous works which revealed
the dynamical nature of pseudospin symmetry. In this work we will
perform a similar calculation, including now a tensor coupling
term, and using again an energy decomposition similar to the one
used in \cite{pmmdm2}. We find that the tensor coupling potential
has a noticeable effect on the pseudospin splittings. We study in
particular detail the changes of the radial wave functions for the
pseudospin partners and the differences between low lying and near
to the Fermi surface pseudospin doublets.

This paper is organized as follows. In Sec.~\ref{SubSec:dirac_hamiltoninan} we
present the Lagrangian for fermion fields coupled to external scalar, vector and
tensor fields and obtain the corresponding single-particle Dirac Hamiltonian. The Dirac
equations of motion are obtained in Sec.~\ref{SubSec:dirac_eq_motion}, with emphasis
on the second-order differential equations for the upper and lower
components of the Dirac spinor. In Sec.~\ref{SubSec:energy_decomp} we perform
the energy decomposition based on the second-order differential equation for
the lower component of the spinor, which will allow us to analyze the contribution
of the radial tensor potential $U$ for the pseudospin energy splittings.
The results of the calculation using mean-fields with Woods-Saxon
shape for $S$ and $V$ radial potentials (thereby also fixing the tensor
potential) are presented in Sec.~\ref{Sec:woods-saxon_results}, together
with a discussion
of the effects of the tensor potential on the neutron pseudospin partners,
both for deep levels and levels near to the Fermi surface. The wavefunctions
for the radial lower components of those levels are also plotted and their
influence on the pseudospin splittings are discussed.
Finally, our conclusions are summarized in Sec.~\ref{conclusions}.


\section{Dirac equation with isoscalar tensor coupling}
\label{Sec:dirac_tensor}


\subsection{Dirac Hamiltonian}
\label{SubSec:dirac_hamiltoninan}

Using the conventions of G.~Mao \cite{Mao}, the nucleon-meson Lagrangian density
of a nuclear mean-field theory with nucleons interacting with $\sigma$, $\omega$
and $\rho$ mesons, in which a tensor (derivative) coupling is
included, reads ($\hbar=c=1$)
\begin{equation}
{\cal
L}=\bar\Psi(i\gamma^\mu\partial_\mu-M)\Psi-g_\sigma\bar\Psi\sigma\Psi
-g_\omega\bar\Psi\gamma_\mu\Psi\omega^\mu-g_\rho
\bar\Psi\gamma_\mu\frac{\vec\tau}2\cdot\Psi\vec\rho^{\,\mu}-
\frac{f_\omega}{4M}\bar\Psi\sigma^{\mu\nu}\Psi\omega_{\mu\nu}-
\frac{f_\rho}{4M}\bar\Psi\sigma^{\mu\nu}\frac{\vec\tau}2\cdot\Psi\vec\rho_{\mu\nu}\
.
\end{equation}
In this Lagrangian,
$\omega_{\mu\nu}=\partial_\mu\omega_\nu-\partial_\nu\omega_\mu$,
$\vec\rho_{\mu\nu}=\partial_\mu\vec\rho_\nu-\partial_\nu\vec\rho_\mu$
and $\sigma^{\mu\nu}=i/2[\gamma^\mu,\gamma^\nu]$. Here the only
vector mesons we are going to consider are the isoscalar $\omega$
mesons. Furthermore, in a mean-field theory, the meson fields are
static and only the time-like component is considered, i.e., we
have
\[
\sigma=\sigma(\bmath r)\qquad\omega_\mu=\omega(\bmath
r)\delta_{\mu\,0}\ .
\]
The Hamiltonian density is then given by
\begin{equation}
{\cal H}=\frac{\partial{\cal L}}{\partial(\partial_0\Psi)}\partial_0\Psi-{\cal L}
=\bar\Psi\gamma^0\partial_0\Psi-{\cal L}=-i\Psi^\dagger\bmath\alpha\ldotp\nabla\Psi+\bar\Psi M\Psi+g_\sigma\bar\Psi\sigma\Psi+g_\omega\Psi^\dagger\Psi\,\omega
-i\frac{f_\omega}{2M}\Psi^\dagger\beta\bmath\alpha\ldotp\nabla\omega\,\Psi
\,.\label{Eq:hamilt_density}
\end{equation}
The corresponding single-particle (Dirac) Hamiltonian is
\[
H=-i\,\bmath\alpha\ldotp\nabla+\beta(
M+g_\sigma\sigma)+g_\omega\omega
-i\,\frac{f_\omega}{2M}\beta\bmath\alpha\ldotp\nabla\omega\ .
\]
If we now define the scalar $S$ and vector $V$ potentials as,
respectively, $S=g_\sigma\sigma$ and $V=g_\omega\omega$, this last
equation reads
\begin{equation}
\label{H1}
H=-i\,\bmath\alpha\ldotp\nabla+\beta(M+S)+V-i\,\beta\,\bmath\alpha\cdot\bmath{\cal
U},
\end{equation}
where $\bmath{\cal U}=f_\omega/(2M)\,\nabla\omega$. Moreover, if
the field $\omega(\bmath r)$ is just a function of the radial
coordinate $r$, this Hamiltonian becomes
\begin{equation}
\label{H2}
H=-i\,\bmath\alpha\ldotp\nabla+\beta(M+S)+V-i\,\beta\,\bmath\alpha\cdot
\hat{\bmath{r}}\, U,
\end{equation}
where $U$ is the radial function
\begin{equation}\label{Eq:U_tensorial}
U(r)=\frac{f_\omega}{2M}\,\omega'=\frac1{2M}\frac{f_\omega}{g_\omega}\;V'\,.
\end{equation}
In the remainder of the paper we will use the notation $f_v\equiv
f_\omega/g_\omega$ for the sake of simplicity.

\subsection{Equations of motion}
\label{SubSec:dirac_eq_motion}

The Dirac equation for nucleons with tensor coupling is written as
\begin{equation}
\label{eq_Dirac}
H\Psi=\mathcal{E}\,\Psi\ ,
\end{equation}
where $H$ given by Eq.~(\ref{H2}).

It is instructive to decompose this equation into two second-order equations
for the upper and lower components of the spinor $\Psi$, but retaining their
spinor structure. To this end, we use the projectors $P_\pm=(I\pm\beta)/2$
applied to $\Psi$, i.e., define the spinors $\Psi_\pm=P_\pm\Psi$.
Applying $P_\pm$ to the left of the Dirac equation (\ref{eq_Dirac}) we get
\begin{eqnarray}
\bmath\alpha\cdot\bmath p\,\Psi_-+(M+S+V)\Psi_+-i\,\bmath\alpha\cdot\bmath{\hat r}\,U\,\Psi_-
=\mathcal{E}\,\Psi_+\\
\bmath\alpha\cdot\bmath p\,\Psi_++(-M-S+V)\Psi_-+i\,\bmath\alpha\cdot\hat{\bmath r}\,U\,\Psi_+
=\mathcal{E}\,\Psi_-\ ,
\end{eqnarray}
or still, defining $\Sigma=V+S$, $\Delta=V-S$, and
$E=\mathcal{E}-M$,
\begin{eqnarray}
\label{eq_psi-}
(\bmath\alpha\cdot\bmath p-i\,\bmath\alpha\cdot\hat{\bmath r}\,U)\Psi_-&=&(E-\Sigma)\Psi_+\\
\label{eq_psi+}
(\bmath\alpha\cdot\bmath p+i\,\bmath\alpha\cdot\hat{\bmath r}\,U)\Psi_+&=&
(E+2M-\Delta)\Psi_-\ .
\end{eqnarray}
Using the formulas in the Appendix we get finally
\begin{eqnarray}
\hspace*{-1.8cm}\bmath p^2\Psi_-+\bigg(U^2-U'-2\,\frac Ur-\frac{\Sigma'U}{E-\Sigma}\bigg)\Psi_-
-\frac{\Sigma'}{E-\Sigma}\frac{\partial\Psi_-}{\partial r}+
\bigg(-4\,U\,&+&2\frac{\Sigma'}{E-\Sigma}\bigg)
\frac{\bmath L\cdot\bmath S}r\Psi_-\nonumber\\
\label{Eq_Psi-} &=&(E-\Sigma)(E+2M-\Delta)\Psi_-
\end{eqnarray}
\begin{eqnarray}
\bmath p^2\Psi_++\bigg(U^2+U'+2\,\frac Ur+\frac{\Delta'U}{E+2M-\Delta}\bigg)\Psi_+
-\frac{\Delta'}{E+2M-\Delta}\frac{\partial\Psi_+}{\partial r}+
\bigg(4\,U\,&+&2\frac{\Delta'}{E+2M-\Delta}\bigg)
\frac{\bmath L\cdot\bmath S}r\Psi_+\nonumber\\
\label{Eq_Psi+} &=&(E-\Sigma)(E+2M-\Delta)\Psi_+ .
\end{eqnarray}
The terms with $\bmath L\cdot\bmath S$ for the upper and lower
components are spin-orbit and pseudospin orbit coupling terms,
respectively. We can note immediately that the tensor potential
contributes to both. A full analysis will be done in the next
section.

If $S$, $V$ and $U$ are radial functions, then the general
solution of Eq.~(\ref{eq_Dirac}) is
\begin{equation}
\Psi_{\kappa m}(\mbox{\boldmath $r$})=\left(
\begin{array}{c}
\displaystyle i\frac{g_{\kappa}(r)}{r} \mathcal{Y}_{\kappa m}(%
\mbox{\boldmath $\hat{r}$}) \\
\noalign{\vskip.2cm} \displaystyle \frac{f_\kappa(r)}{r} \mathcal{Y}%
_{-\kappa m}(\mbox{\boldmath $\hat{r}$})
\end{array}
\right) \, .  \label{gen_spinor}
\end{equation}
Here $\kappa$ is the quantum number related to the total angular
momentum $j$ and orbital momentum $l$ by
\begin{equation}
\label{def_kappa}
\kappa =\left\{
\begin{array}{ccc}
-(l+1) & =-\left( j+1/2\right) , &\quad j=l+1/2\  \\[0.2cm]
l & =+\left( j+1/2\right) , &\quad j=l-1/2\
\end{array}\right. \ .
\end{equation}
The spinor spherical harmonics $\mathcal{Y}_{\kappa m}$ result
from the coupling of the two-dimensional spinors to the
eigenstates of orbital angular momentum and form a complete
orthonormal set. Through the relations
\begin{eqnarray}
j&=&|\kappa|-\frac12\\
\ell&=&|\kappa|+\frac12\bigg(\frac{\kappa}{|\kappa|}-1\bigg)
\end{eqnarray}
one sees that, if the upper component of the spinor in Eq.~(\ref{gen_spinor}) 
has a orbital quantum number $l$, the lower
component (which has quantum number $-\kappa$)
must have a orbital angular momentum $\tilde
l=l-\kappa/|\kappa|$. This quantum number has been associated with
the pseudospin symmetry \cite{gino}.

Using the property $\bmath\sigma \cdot
\hat{\bmath r}\,\mathcal{Y} _{\kappa
m}=-\mathcal{Y}_{-\kappa m}$, Eqs.~(\ref{eq_psi-}) and (\ref{eq_psi+})
reduce to a set of two coupled first-order
ordinary differential equations for the radial upper and lower components $%
g_{\kappa}$ and $f_{\kappa}$, namely,
\begin{eqnarray}
\biggl[\frac{d \ }{d r}+\frac{\kappa}{r}-U(r)\biggr]%
g_{\kappa}(r)&=&[E+2M-\Delta(r)]f_{\kappa}(r)\,,
\label{Eq:D1ordRadup}
\\[0.5cm]
\biggl[\frac{d \ }{d r}-\frac{\kappa}{r}+U(r)\biggr]%
f_{\kappa}(r)&=&-[E-\Sigma(r)]g_{\kappa}(r)\, .
\label{Eq:D1ordRadlow}
\end{eqnarray}

Similarly, from Eqs.~(\ref{Eq_Psi-}) and (\ref{Eq_Psi+}) we arrive at
the following second order differential equations for $g_{\kappa}$ and
$f_{\kappa}$ :
\begin{eqnarray}
\biggl\{\frac{d^{2}\,}{dr^{2}} &-&\frac{\kappa (\kappa +1)%
}{r^{2}}+\frac{\Delta ^{\prime }}{E+2M-\Delta (r)}\biggl[\frac{%
d\,\,}{dr}+\frac{\kappa }{r}-U(r)\biggr]+2\kappa \frac{U(r)%
}{r}-U^{\prime }(r)-U^{2}(r)\biggr\}g_{\kappa }(r)  \nonumber \\
 &=&-[E-\Sigma (r)][E+2M-\Delta (r)]g_{\kappa }(r)\,,
\label{Eq:D2ordgOHGeral}
\end{eqnarray}
\begin{eqnarray}
\biggl\{\frac{d^{2}\,}{dr^{2}} &-&\frac{\kappa (\kappa -1)%
}{r^{2}}+\frac{\Sigma ^{\prime }}{E-\Sigma (r)}\biggl[\frac{%
d\,}{dr}-\frac{\kappa }{r}+U(r)\biggr]+2\kappa \frac{U(r)}{%
r}+U^{\prime }(r)-U^{2}(r)\biggr\}f_{\kappa }(r)  \nonumber \\
&=&-[E-\Sigma (r)][E+2M-\Delta (r)]f_{\kappa }(r).
\label{Eq:D2ordfOHGeral}
\end{eqnarray}
These two equations show explicitly the new terms that depend on $U(r)$ and are originated by the tensor interaction. In particular, the term $2\kappa U(r)/r$, which is the same for the upper and lower component, is the modification in the spin-orbit and pseudospin-orbit terms, respectively, generated by the tensor interaction.


\subsection{Energy decomposition and sum rule}
\label{SubSec:energy_decomp}

The terms in Eq.~(\ref{Eq_Psi-}) with denominator $E-\Sigma$,
which have a singularity in $E=\Sigma$, fulfill a sum rule coming
from Eq.~(\ref{alpha_sigmal}). If one divides each member of that
equation by $2M^*=E+2M-\Delta$, left multiply them by
$\Psi_-^{\dagger}$ and integrate, one gets
\begin{eqnarray}
\hspace*{-.7cm}
\int\Psi_-^{\dagger}\frac{i\,\bmath\alpha\cdot\hat{\bmath r}}{2M^*}\,
\Sigma'\,\Psi_+\,d^3\bmath r&=&
{\rm P}\,\int\Psi_-^{\dagger}\frac{\Sigma'}{E-\Sigma}\frac{1}{2M^*}
\frac{\partial\Psi_-}{\partial r}\,\Psi_-\,d^3\bmath r\nonumber\\
&+&{\rm P}\,\int\Psi_-^{\dagger}\frac{\Sigma'}{E-\Sigma}\frac U{2M^*}\,\Psi_-\,d^3\bmath r
-{\rm P}\,\int\Psi_-^{\dagger}\frac{\Sigma'}{E-\Sigma}\frac{1}{M^*}
\frac{\bmath L\cdot\bmath S}r\,\Psi_-\,d^3\bmath r\ ,
\end{eqnarray}
where P stands for the principal value of the integral. In
terms of the radial functions $g_\kappa$ and $f_\kappa$ the sum
rule reads
\begin{equation}
-\int_0^\infty f_\kappa\frac{\Sigma'}{2M^*}\,g_\kappa\,dr=
{\rm P}\int_0^\infty\frac{\Sigma'}{E-\Sigma}\frac{1}{2M^*}\,f_\kappa
\bigg(\frac{f_\kappa}r\bigg)'\,r\,dr
+{\rm P}\int_0^\infty\frac{\Sigma'}{E-\Sigma}\frac{U}{2M^*}\,f_\kappa^2
\,dr
+{\rm P}\int_0^\infty\frac{\Sigma'}{E-\Sigma}\frac{1-\kappa}{2M^*}\,f_\kappa^2
\,dr\ .
\end{equation}
This sum rule can be used to check the numerical results.

The energy decomposition of Eq.~(\ref{Eq_Psi-}) can be performed by dividing it by
$2M^*$ and computing its expectation value for the spinor $\Psi_-$, yielding
\vskip-0.2cm
\begin{equation}
\label{energ_decomp}
\langle\frac{\bmath p^2}{2M^*}\rangle+\langle V_U\rangle+\langle V_{\Sigma'U}\rangle+
\langle V_{\rm Darwin}\rangle +\langle V_{\rm PSO}\rangle+\langle\Sigma\rangle=E ,
\end{equation}
\vskip-0.3cm
where
\begin{eqnarray}
V_U&=&\frac1{2M^*}\bigg(U^2-U'-2\,\frac Ur\bigg)\nonumber\\
V_{\Sigma'U}&=&-\frac1{2M^*}\frac{\Sigma'U}{E-\Sigma}\nonumber\\
V_{\rm Darwin}&=&-\frac1{2M^*}\frac{\Sigma'}{E-\Sigma}\frac{\partial\hfill}{\partial r}\nonumber\\
V_{\rm PSO}&=&\frac1{M^*}\bigg(-2\,U\,+\frac{\Sigma'}{E-\Sigma}\bigg)\frac{\bmath L\cdot\bmath S}r\label{V_PSO}\\
\noalign{\goodbreak}
\langle {\cal O} \rangle&\equiv&\frac{\ds\int\Psi_-^{\dagger}{\cal
O}\,\Psi_-\,d^3\bmath r}
{\ds\int\Psi_-^{\dagger}\,\Psi_-\,d^3\bmath r}\ .\nonumber
\end{eqnarray}
\vskip-0.3cm
For the terms with $E-\Sigma$ in denominator the integral is taken
in the principal value sense.


\section{Tensor coupling with mean-field Woods-Saxon potentials}
\label{Sec:woods-saxon_results}

As stated above, the aim of this paper is to study the effect of
the tensor coupling on the pseudospin splitting in nuclei.
In previous works, in which we
studied pseudospin symmetry in nuclei \cite{pmmdm1}, we solved numerically
the Dirac equation with central mean-field potentials with Woods-Saxon shapes.
Although these potentials are not full self-consistent relativistic potentials
derived from meson fields, they are realistic enough to
be applied to many nuclei. In this paper we follow the same approach,
namely we consider the sum and difference potentials
$\Sigma$ and $\Delta$ to be of the general form
\begin{equation}
P(r)=\frac{P_0}{{1+\exp [(r-R)/a]}}\ ,
\label{WSaxon}
\end{equation}
whereas the tensor potential $U(r)$ is obtained by
\begin{equation}
\label{def_U}
U(r)=\frac{f_v}{2M}\;V'=
\frac {f_v}{2M}\;\frac{\Sigma'+\Delta'}2\ .
\end{equation}

The depth, $P_0$, the radius (range), $R$, and the diffusivity,
$a$, for $\Sigma$ and $\Delta$ are fitted to reproduce the
single-particle spectrum of $^{208}$Pb \cite{pmmdm1,pmmdm2}.

\begin{figure}[!ht]
\begin{center}
\includegraphics[width=11cm]{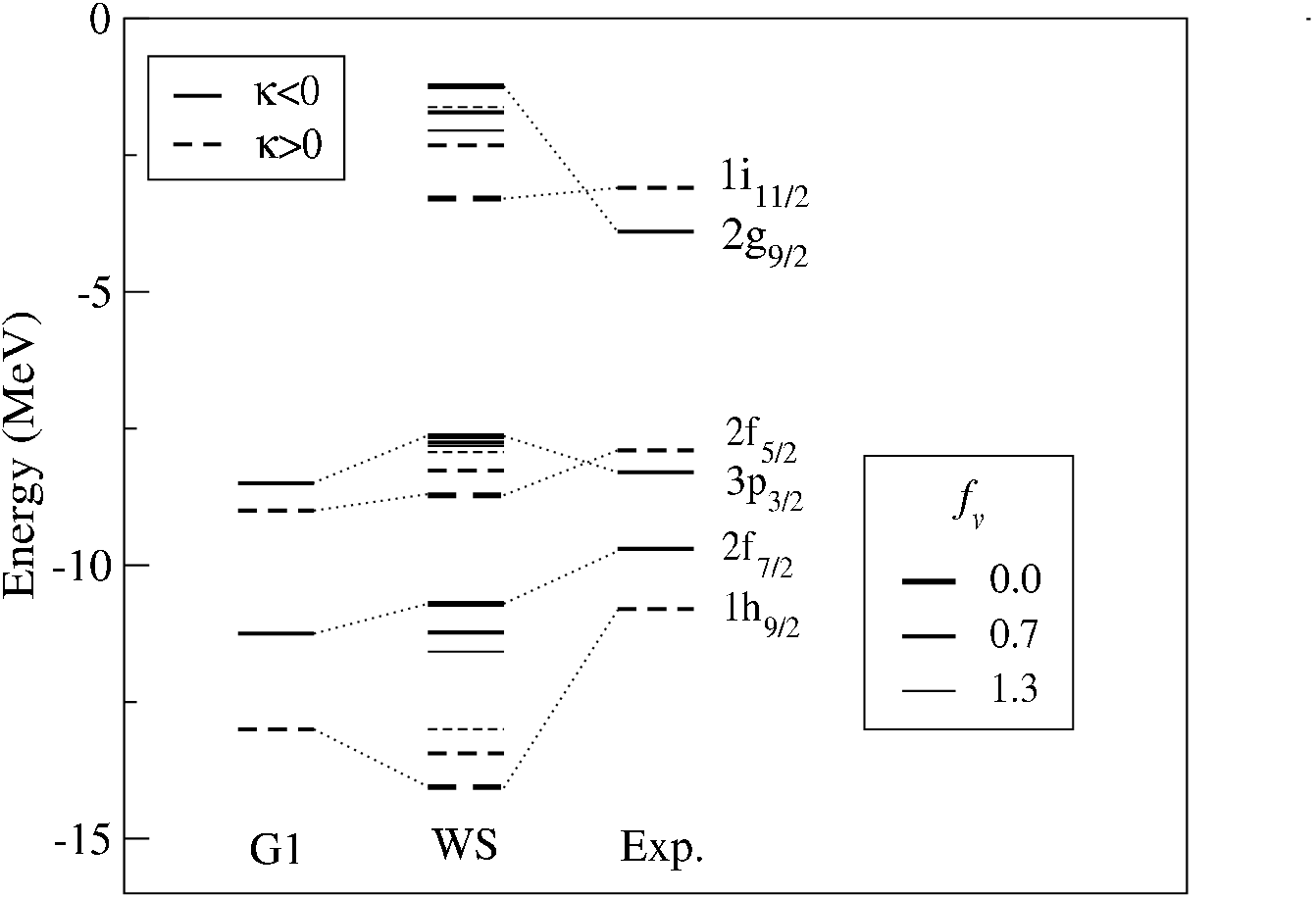}
\end{center}
\caption{Calculated neutron single-particle energy levels of the
pseudospin partners $[2f_{7/2}-1h_{9/2}]$, $[2f_{5/2}-3p_{3/2}]$
and $[1i_{11/2}-2g_{9/2}]$ in $^{208}$Pb. The left-most
values are the corresponding values of the model G1 of
Furnsthal {\it et al.} \cite{furnstahl1} and the experimental values \cite{campi}
are in the right-most column. In the middle
column (WS) is our calculation with Woods-Saxon potentials for three values of 
the tensor coupling strength $f_v$,
corresponding to lines of different thickness. Pseudospin partners levels with $\kappa<0$
are represented by full lines whereas those with $\kappa >0$ have dashed lines.
The Woods-Saxon parameters used to fit $^{208}$Pb neutron energy levels
are $R$ = 7 fm,
$\Delta_0=650$ MeV, $\Sigma_0=-66$ MeV and $a=0.6$ fm. \hfill\ }
\label{Fig:fig1}
\end{figure}

Using the general Woods-Saxon form in Eq.~(\ref{WSaxon}) for $\Sigma$ and
$\Delta$, and with $U$ given by Eq.~(\ref{def_U}), we solved
numerically the coupled first-order Dirac equations
(\ref{Eq:D1ordRadup}) and (\ref{Eq:D1ordRadlow}).
The single-particle energy levels are shown in Fig.~1, when $f_v$ varies from 0 to
1.3. This range of values is consistent to that one found in
Ref.~\cite{furnstahl} for fittings of RMF and point coupling models to
nucleon observables.

One sees clearly from Fig.~1 that turning on the tensor coupling
decreases the pseudospin splittings of the levels near the Fermi
surface. For $f_v=1.3$ the pair $[2f_{5/2}-3p_{3/2}]$ becomes
almost degenerate while the $[1i_{11/2}-2g_{9/2}]$ doublet even
reverses its order. Interestingly, the experimental energy values
for these two pairs show this order reversal, which is not
reproduced by the model calculations without tensor coupling. In
Fig.~\ref{Fig:fig1}, when we increase the tensor coupling $f_v$,
the energy for pseudospin partners with $\kappa < 0$ become deeper
and those with $\kappa > 0$  become more unbounded. This systematics
indicates that pseudospin symmetry is improved by the tensor
interaction.

In order to understand better why and how tensor coupling affects
pseudospin splittings, we computed the contributions from the
terms of the energy decomposition in Eq.~(\ref{energ_decomp}) to
the energy splittings for both low-lying and close to the Fermi
surface pseudospin partners.

\begin{table}
\setlength{\tabcolsep}{.15cm}
\begin{center}
\begin{tabular}{||c|r|r|r|r|r|r|r|r||}
\hline\hline
partners    &\omit\hfill$f_v$\hfill\vline  &  $\langle p^2/2M^{*}\rangle$
&\multicolumn{1}{c|}{$\langle V_U\rangle$} &\multicolumn{1}{c|}{$\langle V_{\Sigma'U}\rangle$}
&  $\langle V_{\rm Darwin}\rangle$     &\multicolumn{1}{c|}{$\langle V_{\rm PSO}\rangle$}   &\multicolumn{1}{c|}{$\langle \Sigma\rangle$}     &\multicolumn{1}{c||}{$E$}\\%
\hline\hline
%
                  & 0.0 &  24.4396 & 0.0000  &  0.0000  &  -3.9527  & -0.5852  &  -61.4644  & -41.5627\\%
\rb{$2s_{1/2 }$}  & 1.3 &  23.9037 & 0.2351  & -0.0943  &  -3.6870  & -0.5632  &  -61.9114  & -42.1170\\%
\hline
                  & 0.0 &  21.1032 & 0.0000  &  0.0000  &  -0.8106  &  0.0966  &  -64.4159  & -44.0266\\%
\rb{$1d_{3/2 }$}  & 1.3 &  20.7075 & 0.1459  & -0.0712  &  -0.4170  &  0.1559  &  -64.6181  & -44.1170\\%
\hline\noalign{\vspace*{.7pt}}\hline
                  & 0.0 &  33.2950 & 0.0000  & 0.0000   &  -2.7538  & -1.6340  & -60.2687   & -31.3615\\%
\rb{$2p_{3/2 }$}  & 1.3 &  32.7195 & 0.2465  &  0.1223  &  -2.7319  & -1.5835  &  -60.8762  & -32.1033\\%
\hline
                  & 0.0 &  28.5114 & 0.0000  &  0.0000  &   0.7303  &  0.5384  &  -64.5165  & -34.7365\\%
\rb{$1f_{5/2 }$}  & 1.3 &  28.2218 & 0.2598  & -0.0021  &   0.9522  &  0.6780  &  -64.6667  & -34.5571\\%
\hline\noalign{\vspace*{.7pt}}\hline
                  & 0.0 &  52.9135 & 0.0000  &  0.0000  &   0.8482  & -1.6013  &  -59.8025  &  -7.6420\\%
\rb{$3p_{3/2 }$}  & 1.3 &  52.6913 & 0.5000  &  0.1669  &   0.9638  & -1.9231  &  -60.2278  &  -7.8280\\%
\hline
                  & 0.0 &  50.1727 & 0.0000  &  0.0000  &   2.3696  &  1.0254  &  -62.2874  &  -8.7197\\%
\rb{$2f_{5/2 }$}  & 1.3 &  50.0072 & 0.6256  &  0.1427  &   2.3831  &  1.3089  &  -62.3270  &  -7.8595\\%
\hline\noalign{\vspace*{.7pt}}\hline
                  & 0.0 &  55.7666 & 0.0000  &  0.0000  &   2.8295  & -6.4555  &  -53.3816  &  -1.2410\\%
\rb{$2g_{9/2 }$}  & 1.3 &  56.0706 & 0.5622  &  0.2855  &   2.6487  & -7.2110  &  -54.4705  &  -2.1140\\%
\hline
                  & 0.0 &  51.2033 & 0.0000  &  0.0000  &   3.1788  &  3.3530  &  -61.0308  &  -3.2958\\%
\rb{$1i_{11/2}$}  & 1.3 &  51.6221 & 0.8997  &  0.1529  &   2.9268  &  4.1391  &  -61.2232  &  -1.4826\\%
\hline
\end{tabular}
\end{center}
\caption{Values of energies and terms in the decomposition (\ref{energ_decomp}) for the
pseudospin partners $[2s_{1/2} - 1d_{3/2}]$, $[2p_{3/2} - 1f_{5/2}]$, $[3p_{3/2} - 2f_{5/2}]$ and $[2g_{9/2} - 1i_{11/2}]$ for two values of $f_v$. The energies and expectation values are given in MeV.\hfill\ }
\label{tab1}
\end{table}

In Table~\ref{tab1} we can see what is the contribution of all the terms of that energy decomposition for
the two lowest neutron pseudospin partners ($[2s_{1/2} - 1d_{3/2}]$ and $[2p_{3/2} - 1f_{5/2}]$) and
for the two topmost neutron pseudospin partners ($[3p_{3/2} - 2f_{5/2}]$ and $[2g_{9/2} - 1i_{11/2}]$) for two values of the tensor coupling strength: $f_v = 0.0$ (no tensor coupling) and $f_v=1.3$.
As expected, since tensor interaction is an higher order interaction in the
Lagrangian (derivative term) scaled by $1/M$, the changes in the energy
produced by the potential terms $V_U$ and $V_{\Sigma 'U}$
are small in comparison with the kinetic and potential terms, as we can see from Tab.~\ref{tab1}.
This table also shows that these terms, together with the pseudospin-orbit term, are significantly
bigger for the surface levels than for the lower levels. This agrees with our expectations, referred before, that the effect of the tensor coupling
is larger for the surface levels, since the potential $U$ is
proportional to the derivative of the vector potential. On the other hand, changes induced
by the tensor coupling in
$\langle V_{\Sigma'U}\rangle$ and $\langle V_{\rm Darwin}\rangle$ for surface levels
are smaller.

The smallness of the terms containing the tensor potential in
regard to the kinetic and $\Sigma$ potential terms is misleading
concerning the effect in the pseudospin symmetry, since the
changes of these last terms with $f_v$ are small, whereas the
corresponding changes in $\langle V_{\rm PSO}\rangle$ and $\langle
V_U\rangle$ can be significant for the surface levels, especially
when compared with the energies of these levels. Note that the
values of $\langle V_{\rm PSO}\rangle$ can have quite different
values for different levels. These is due to the respective values
of the $\kappa$ quantum number, as explained below.

\begin{figure}[!ht]
\parbox[!t]{9.5cm}{
\begin{center}
\includegraphics[width=9.5cm,height=7cm]{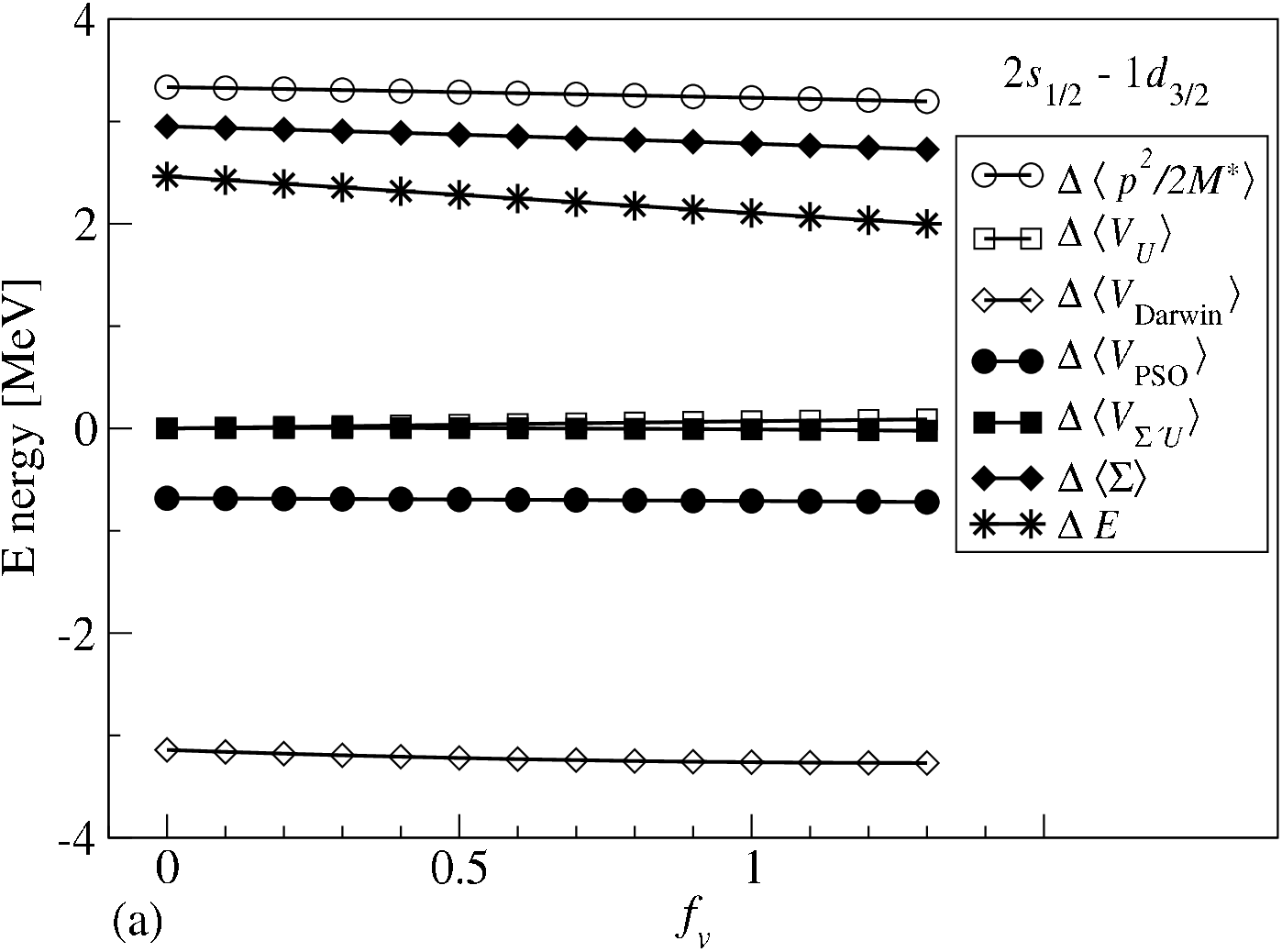}
\end{center}\vspace*{.2cm}}
\parbox[!h]{9.5cm}{
\begin{center}
\includegraphics[width=9.5cm,height=7cm]{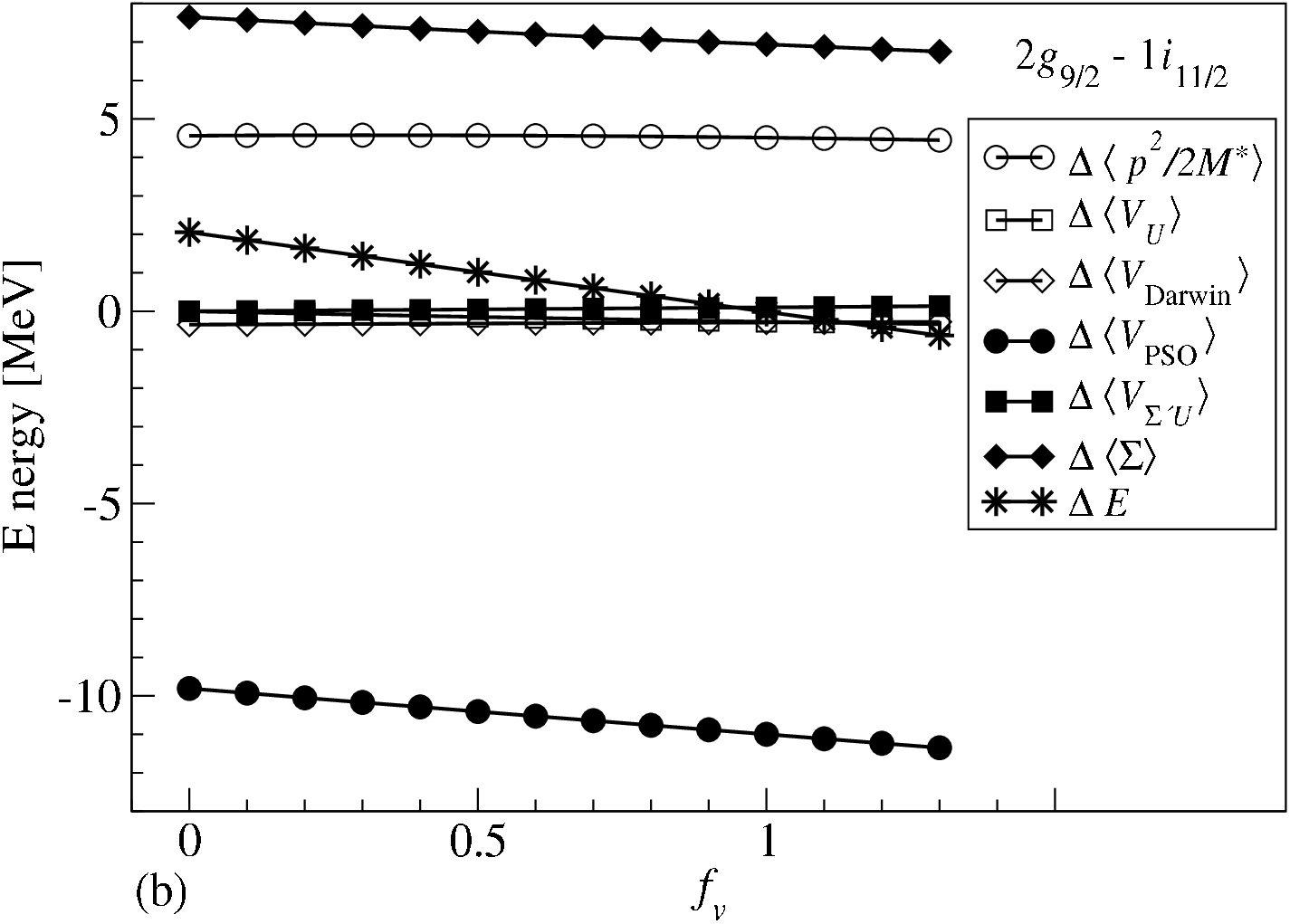}
\end{center}\vspace*{.1cm}}
\caption{The contributions from the terms of the energy
decomposition (\ref{energ_decomp}) for pseudospin doublets (a)
$[2s_{1/2}-1d{3/2}]$ and (b) $[2g_{9/2}-1i_{11/2}]$ when $f_v$
varies from $0$ to $1.3$.\hfill\ } \label{Fig:fig2}
\end{figure}

To have a better understanding of the influence of all these terms in
the pseudospin splittings, we plot in Figs.~\ref{Fig:fig2}(a) and \ref{Fig:fig2}(b)
the splittings and the \textit{differences} of the terms in (\ref{energ_decomp}) for the
pseudospin doublets $[2s_{1/2}-1d_{3/2}]$ and $ [2g_{9/2}-1i_{11/2}]$
as a function of $f_v$.
From these figures is clear that the decrease of the pseudospin energy splitting $\Delta E$
is much more pronounced for the surface doublet than for the
low-lying one, and would be even more if we considered the relative
energy variations.

A more detailed analysis of the several contributions to the pseudospin energy
splittings reveals that for the deep pseudospin doublet the contribution
of the pseudospin-orbit potential $V_{\rm PSO}$ almost does not change with $f_v$,
whereas the contribution of terms like $\langle p^2/2M^*\rangle$ and
$\langle \Sigma\rangle$, which do not depend explicitly on the tensor potential
$U$, is greater. This means that the main contribution to the change of
pseudospin splitting with the strength of the tensor potential comes mainly
via the change of the wave function induced by $U$. Furthermore, the energy splitting
results for the most part from a cancellation of the $\Delta\langle\Sigma\rangle$ and
$\Delta\langle V_{\rm Darwin}\rangle$ contributions and also from
$\Delta\langle p^2/2M^*\rangle$, the contribution from $V_{\rm PSO}$ having
a lesser role. This agrees with previous findings of similar studies of pseudospin
splittings \cite{pmmdm2,prc67}.
\begin{figure}[!ht]
\parbox[!t]{9.5cm}{
\begin{center}
\includegraphics[width=9.5cm,height=7cm]{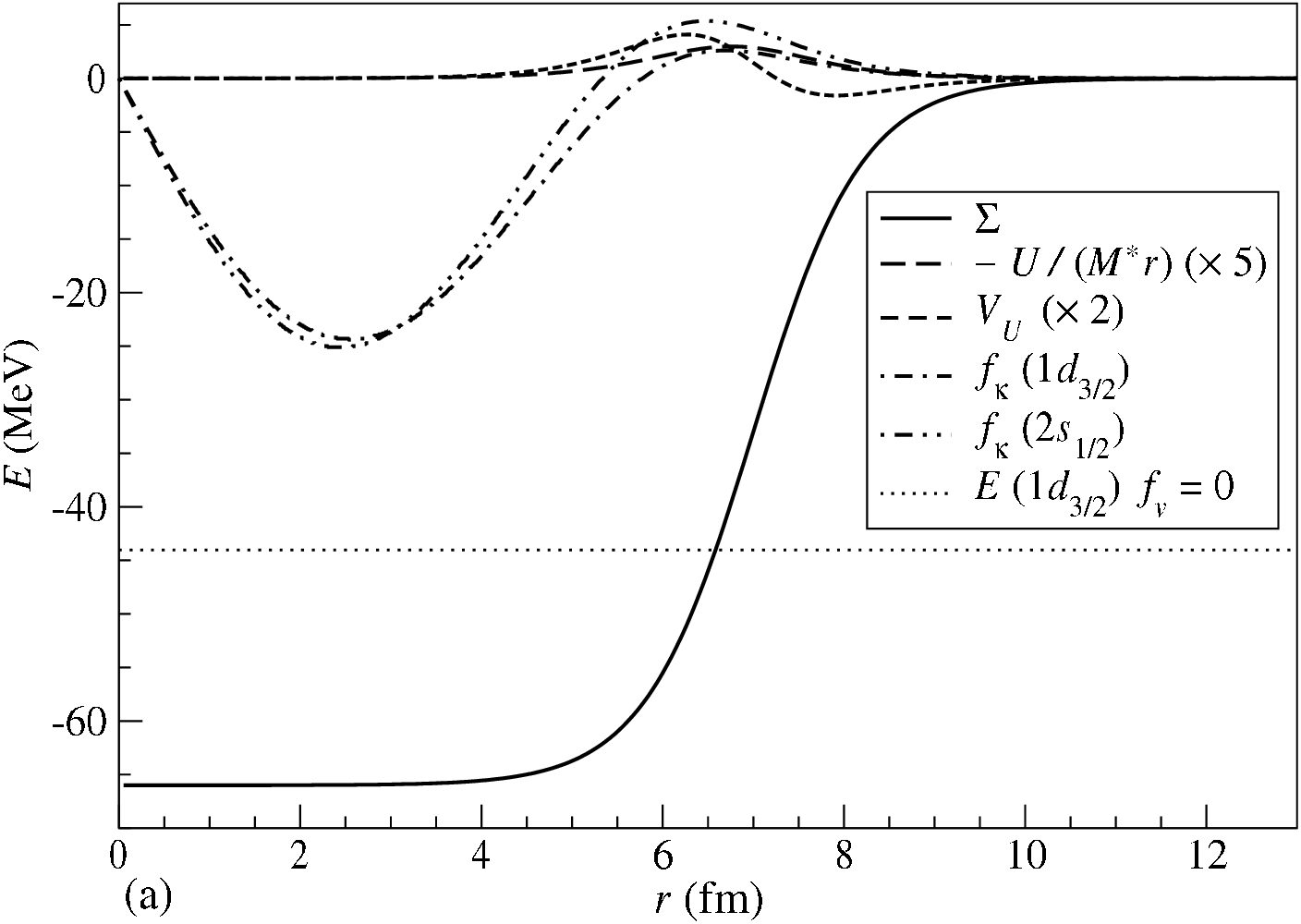}
\end{center}\vspace*{.2cm}}
\hfill
\parbox[!h]{9.5cm}{
\begin{center}
\includegraphics[width=9.5cm,height=7cm]{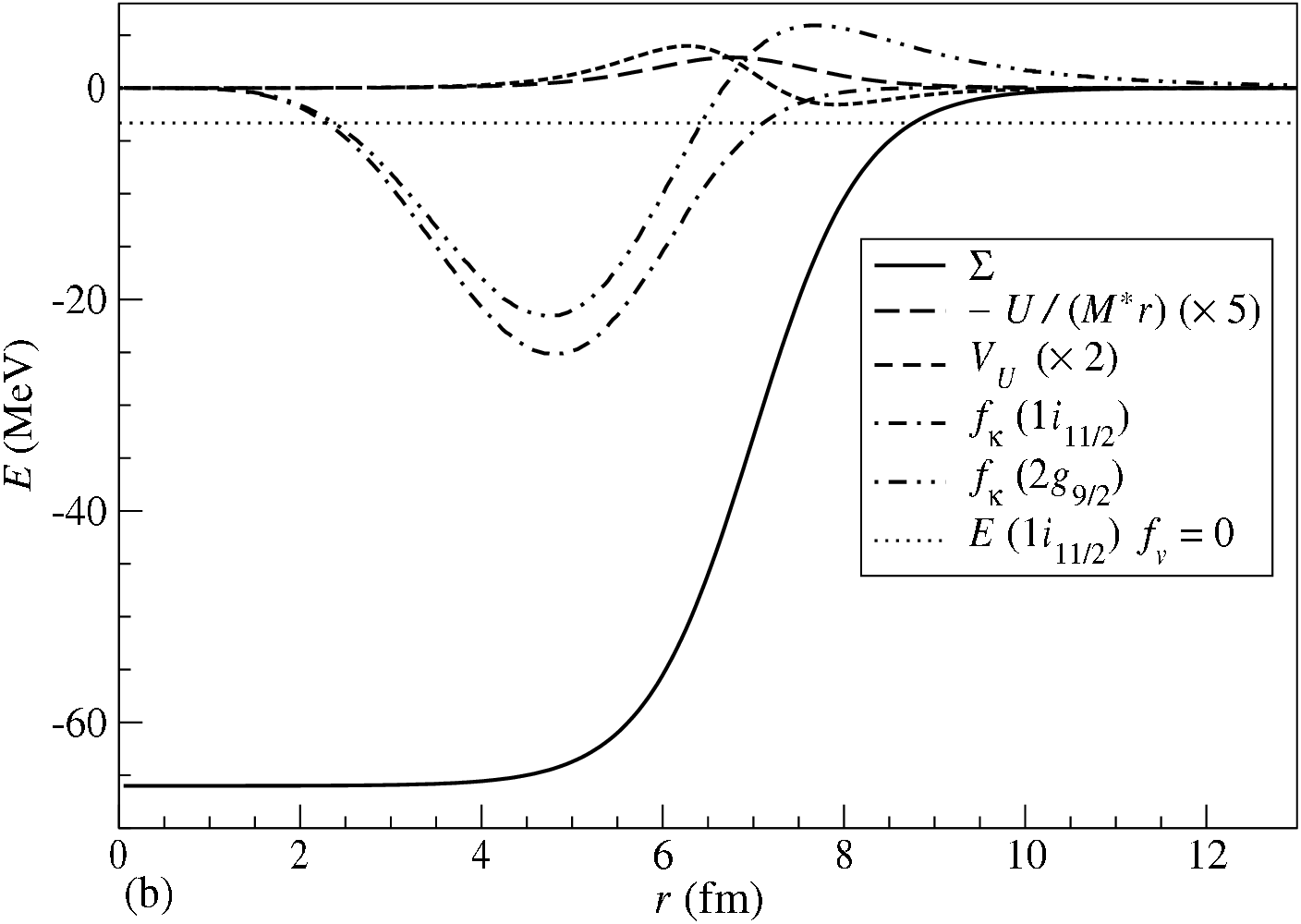}
\end{center}\vspace*{.2cm}
}%
\caption{The potentials $\Sigma$, $V_U$, $-U/(M^*\,r)$, and the
lower radial wave functions $f_\kappa$ for the pseudospin partners
(a) $[2s_{1/2}-1d{3/2}]$ and (b) $[2g_{9/2}-1i_{11/2}]$ when
$f_v=1.2$. The wave functions were normalized such that
they could be plotted side by side with the potentials.\hfill\ } \label{Fig:fig3}
\end{figure}

As far as the upper doublet $[2g_{9/2}-1i_{11/2}]$ is concerned,
much of the previous analysis still holds, except for the fact
that the pseudospin-orbit potential is much stronger and also
changes sensibly as $f_v$ changes, being responsible for most of
the pseudospin splitting. The reason for this, and in particular
the fact that for values of $f_v$ greater than $0.95$ the
splitting becomes negative, i.e., there is a level inversion, must
be found in a more detailed analysis of the contribution of the
tensor potential $U$ for the $V_{\rm PSO}$ potential. From
Eq.~(\ref{V_PSO}) we see that, while $U$ gives a positive
contribution to this potential (since $V'<0$, so that,
from Eq.~(\ref{Eq:U_tensorial}), $-U>0$), 
the effective contribution depends on the
sign of $\Sigma'/(E-\Sigma)\times\bmath L\cdot\bmath S/r$. Since
it changes sign when $r>r_s$, in which $r_s$ is the radius for
which $\Sigma=E$, the net change of $\langle V_{\rm PSO}\rangle$
depends on how the wave function $f_\kappa(r)$ behaves near $r_s$
for both pseudospin partners. What we found is that for these
surface levels, the $U$ contribution for the pseudospin-orbit
potential
\begin{equation}
V_{\rm PSO}^U=-\frac2{M^*}\,U\,\frac{\bmath L\cdot\bmath S}r
\label{V_PSO_U}
\end{equation}
is the dominant one. The reason why the contribution
of this potential to $\Delta V_{\rm PSO}$ for the pair 
$[2g_{9/2}-1i_{11/2}]$ is so big lies
in the fact that it is proportional to $\langle 2\bmath
L\cdot\bmath S\rangle=-(1-\kappa)$, which is equal to $-6$ and $5$
for $2g_{9/2}$ and $1i_{11/2}$ respectively. Since $U$ is
negative, $\langle V^U_{\rm PSO}\rangle (2g_{9/2})- \langle
V^U_{\rm PSO}\rangle (1i_{11/2})$ is negative, and more so as
$f_v$ increases, (see Fig.~\ref{Fig:fig2}(b)), thus explaining
why pseudospin splitting decreases with the increase of $f_v$.

Note that, from Table~\ref{tab1}, one is able to see directly the different effects that
the tensor coupling has on
$\Delta\langle V_{\rm PSO}\rangle$ in a low-lying ($[2p_{3/2} - 1f_{5/2}]$) and a surface
($[3p_{3/2} - 2f_{5/2}]$) pseudospin partner
that have the \textit{same} $\kappa$ values. The values of $\langle V_{\rm PSO}\rangle$ and their
differences change significantly with $f_v$ for the surface partner.

In Figs. \ref{Fig:fig3}(a) and \ref{Fig:fig3}(b) are represented
$\Sigma$, $V_U$, the radial part of the $V_{\rm PSO}^U$ potential
[Eq.~(\ref{V_PSO_U})], given by $-U/(M^*\,r)$, which is always
positive as discussed before, and the lower radial wavefunctions
$f_\kappa$ for the pseudospin partners $[2s_{1/2}-1d_{3/2}]$ and
$[2g_{9/2}-1i_{11/2}]$ respectively, when $f_v=1.2$. Also plotted
is the energy of the lowest lying level of each doublet for
$f_v=0$, allowing to have a rough estimate of the value $r_s$
mentioned above by the intersection of its horizontal line with
the $\Sigma$ potential curve. From these figures one sees clearly
why the tensor potential $U$ has a much larger effect for a
surface pseudospin doublet. Indeed, the lower radial wavefunctions
for these levels have a significant strength near the nucleus
surface, such that $\langle V^U_{\rm PSO}\rangle$ can have a
significant value and have a sizeable contribution to the
pseudospin splitting.


\section{Conclusions}
\label{conclusions} In this paper we have assessed the importance
of the isoscalar tensor coupling to pseudospin symmetry as is
realized in heavy nuclei, using as an example the neutron
pseudospin partners of $^{208}$Pb, calculated within a
relativistic theory with scalar and vector mean-fields
parametrized with Woods-Saxon potential forms. By looking into the
second-order equation for the lower component of the Dirac spinor,
we obtained an expression for the pseudospin-orbit potential,
showing its explicit dependence on the radial tensor potential.
Since this potential comes from a derivative coupling (higher
order term in the Lagrangian) the contribution for the energy from
the potentials originated by that interaction are all scaled by
$1/M$ ($M$ is the nucleon mass) and, because of
that, is very small in comparison with kinetic and potential terms. However, since the spin-orbit interaction (and pseudospin) is a term of the same order, the effect of the
tensor coupling can be significant in this case. In fact, we
conclude that the contribution from $U$ to the pseudospin
potential $V_{\rm PSO}$ is the dominant one for the surface levels.

We have shown that the surface pseudospin partners were
the most affected by the tensor coupling, as was expected, and that this coupling
reduces pseudospin splitting. This reduction can be significative to the point of
inverting the level order, with the states with aligned spin ($j=l+1/2$) having higher
energy than the states with anti-aligned spin.

By analyzing the several contributions to the pseudospin energy splittings, we
were able to confirm a conclusion reached in previous works, namely that the
pseudospin symmetry is realized dynamically in nuclei, resulting from a cancellation
of the various contributions, rather than just the one from the pseudospin-orbit potential. We also found a systematic change in the energy for the pseudospin partners when we increase the tensor coupling $f_v$: states with aligned spin ($\kappa<0$) become deeper in opposite to anti-aligned ones that become more unbounded. This systematics allows us to conclude that pseudospin symmetry is improved by the tensor interaction.
Finally, we were able to show for a surface pseudospin doublet the pseudospin-orbit potential, especially through its tensor potential part, gives a significant contribution for the change of the energy
splitting. This finding is compatible with the dynamical character of the pseudospin symmetry.


\begin{acknowledgments}
We acknowledge financial support from CNPq, FAPESP, and
FCT (POCTI) scientific program. R.L. and M.M acknowledge, in
particular, the CNPq support and A.S.C was also supported by
FAPESP.
\end{acknowledgments}

\appendix*
\section{}

In this Appendix we present the derivation of equations (\ref{Eq_Psi-}) and (\ref{Eq_Psi+}).

\vskip.2cm

If we define the operator
${\cal O}=\bmath\alpha\cdot\bmath p+i\,\bmath\alpha\cdot\hat{\bmath r}\,U$, the second-order
differential equations are obtained by applying ${\cal O}{\cal O}^\dagger$ and ${\cal O}^\dagger{\cal O}$
to $\Psi_-$ and $\Psi_+$, respectively. For $\Psi_-$ we have
\begin{eqnarray}
{\cal O}{\cal O}^\dagger\Psi_-&=&{\cal O}(E-\Sigma)\Psi_+=
[\bmath\alpha\cdot\bmath p\,(E-\Sigma)]\Psi_++ (E-\Sigma){\cal
O}\Psi_+\nonumber\\
\label{OOd-1}
&=&i\,\bmath\alpha\cdot\hat{\bmath r}\,\Sigma'\,\Psi_++(E-\Sigma)(E+2M-\Delta)\Psi_-\nonumber\\
\label{OOd-2} &=&\bigg(\bmath p^2+U^2-U'-2\,\frac
Ur-4\,U\,\frac{\bmath L\cdot\bmath S}r\bigg)\Psi_-\ ,
\end{eqnarray}
where it was used the fact that $\Sigma$, $\Delta$ and $U$ are
radial potentials, and primes denote derivatives with respect to
$r$. $\bmath S$ stands for the $4\times 4$ spin matrix. On the
other hand, from the equation ${\cal
O}^\dagger\Psi_-=(E-\Sigma)\Psi_+$ we obtain
\begin{eqnarray}
{\cal O}^\dagger\Psi_-&=&(\bmath\alpha\cdot\bmath p-i\,\bmath\alpha\cdot\hat{\bmath r}\,U)
\Psi_-\nonumber\\
&=&\bmath\alpha\cdot\hat{\bmath r}(\hat{\bmath r}\cdot\bmath p+
i\,\hat{\bmath r}\times\bmath p\cdot\bmath\Sigma-i\,U)\Psi_-\nonumber\\
&=&-i\,\bmath\alpha\cdot\hat{\bmath r}\bigg(\frac{\partial\Psi_-}{\partial r}+U\,\Psi_-
-2\frac{\bmath L\cdot\bmath S}r\Psi_-\bigg)\nonumber\\
&=&(E-\Sigma)\Psi_+\nonumber\ ,
\end{eqnarray}
which allows us to write
\begin{equation}
\label{alpha_sigmal}
i\,\bmath\alpha\cdot\hat{\bmath r}\,\Sigma'\,\Psi_+=i\,\bmath\alpha\cdot\hat{\bmath r}\,\Sigma'\,
\frac{{\cal O}^\dagger\Psi_-}{E-\Sigma}=
\frac{\Sigma'}{E-\Sigma}\bigg(\frac{\partial\Psi_-}{\partial r}+U\,\Psi_-
-2\frac{\bmath L\cdot\bmath S}r\Psi_-\bigg)\ .
\end{equation}
In the same way, for $\Psi_+$ we have
\begin{eqnarray}
\label{OdO+1}
{\cal O}^\dagger{\cal O}\Psi_+&=&
i\,\bmath\alpha\cdot\hat{\bmath r}\,\Delta'\,\Psi_-+(E-\Sigma)(E+2M-\Delta)\Psi_+\nonumber\\
\label{OdO+2}
&=&\bigg(\bmath p^2+U^2+U'+2\,\frac Ur+
4\,U\,\frac{\bmath L\cdot\bmath S}r\bigg)\Psi_+\ ,
\end{eqnarray}
and, using equation ${\cal O}\Psi_+=(E+2M-\Delta)\Psi_-$, we get
\begin{equation}
\label{alpha_deltal}
i\,\bmath\alpha\cdot\hat{\bmath r}\,\Delta'\,\Psi_-=
\frac{\Delta'}{E+2M-\Delta}\bigg(\frac{\partial\Psi_+}{\partial r}-U\,\Psi_+
-2\frac{\bmath L\cdot\bmath S}r\Psi_+\bigg)\ .
\end{equation}

Using the equations (\ref{OOd-1}) and (\ref{alpha_sigmal}) for
$\Psi_-$ and (\ref{OdO+1}) and (\ref{alpha_deltal}) for $\Psi_+$,
we get finally the equations (\ref{Eq_Psi-}) and (\ref{Eq_Psi+}).



\end{document}